\documentclass[%
 reprint,
 amsmath,amssymb,
 aps,
]{revtex4-1}
\usepackage{physics}
\usepackage{setspace}
\usepackage[flushleft]{threeparttable}
\usepackage{amsmath}
\usepackage{mathtools}
\usepackage{graphicx}
\usepackage{dcolumn}
\usepackage{bm}
\usepackage{ulem}
\usepackage{footnote}

\begin{document}

\preprint{APS}

\title{Multi-qubit production in spontaneous parametric down conversion}

\author{Phillip Heitert$^{*}$, Finn Buldt$^{*}$, Pascal Bassene, Moussa N'Gom$^{\dagger}$}
 
\affiliation{%
Department of Physics, Rensselaer Polytechnic Institute\\ 110 8th St., Troy, NY, 12180, USA \\
$^*$ These authors contributed equally to this work. \\
$^\dagger$ Corresponding email: ngomm@rpi.edu (alt: mngom@umich.edu) 
}%

\date{\today}
          
\begin{abstract}
We present a novel yet simple approach to produced multiple entangled photon pairs through spontaneous parametric downconversion. We have developed Gaussian masks to subdivide the pump beam before passing it through a nonlinear medium. In this way, we are able to observe simultaneous separate down-converted emission cones with spatial overlap.  The technique we employ can be used to greatly increase the dimensionality of entangled photonic systems generated from spontaneous parametric down conversion, affording greater scalability to optical quantum computing than previously explored. 
\end{abstract}

\maketitle

\section{\label{sec:level1}Introduction}

 An important process in the field of quantum optics is spontaneous parametric down conversion (SPDC), a phenomenon wherein excitations within a nonlinear crystal are used to produce correlated photon pairs. To effectively produce more than two entangled photons through SPDC, it is common to employ multiple nonlinear optical elements or UV pulsed lasers. The latter are inconvenient due to very low probabilities for simultaneous conversion processes, and the former requires tedious optical alignment \cite{GHZExp, QOBook}. Although a well-studied phenomena, we present a novel yet simple approach to increasing the entanglement products of such processes. Furthermore, the versatility of such an approach lends itself to increasing the possible configurations available to create photons pairs with desired correlations in an optics environment (i.e. polarization, wavelength , etc.). 

\begin{figure}
\centering\includegraphics[width=\columnwidth]{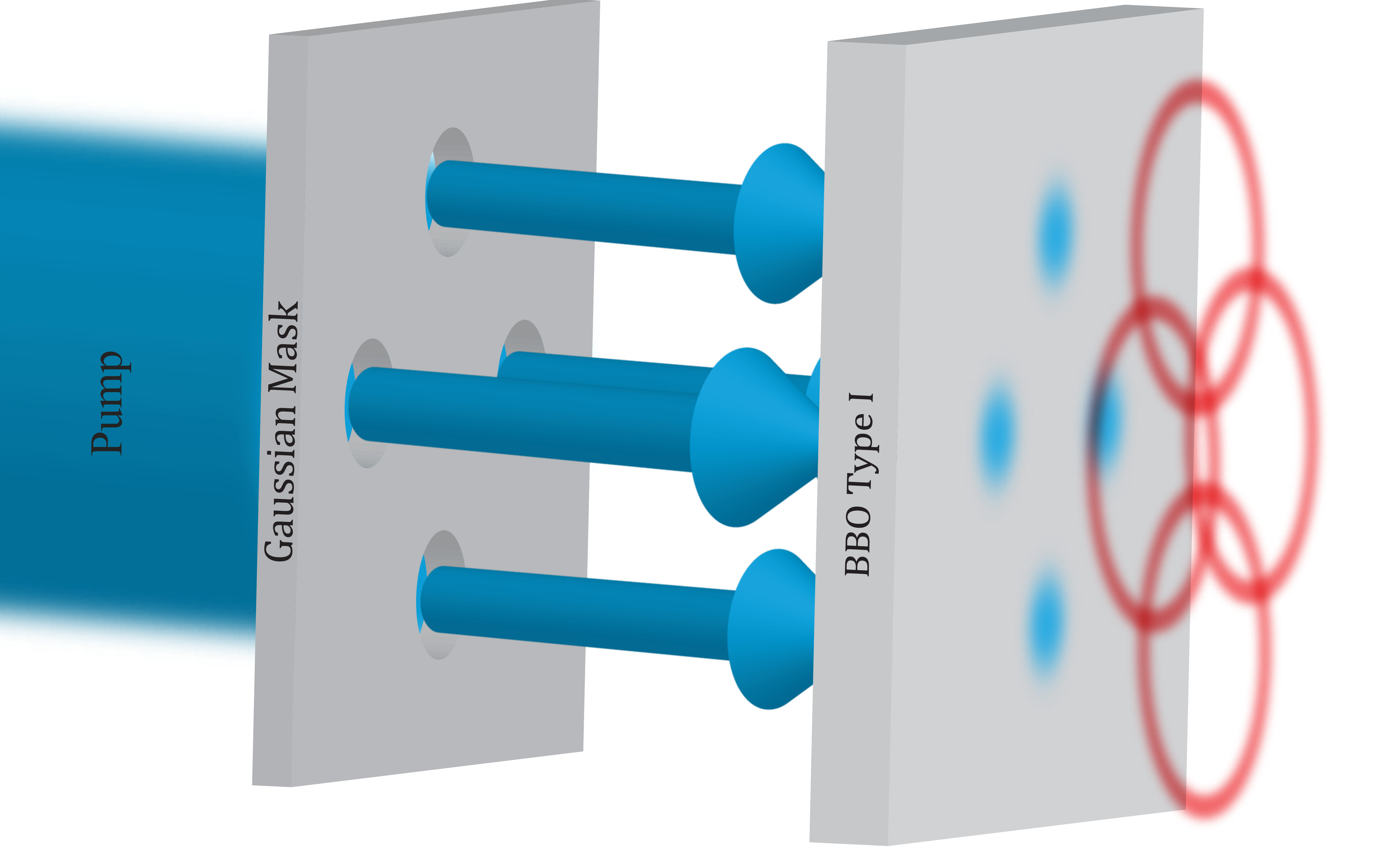}
\caption{A 4-aperture Gaussian mask (GM) is illustrated. Each `sub-Gaussian' input generates its own SPDC emission cone. The GM design controls the number and the manner of overlap of the down-converted cones}
\label{Mask Figure}
\end{figure}

 We conduct degenerate SPDC utilizing both type I and type II beta-Barium Borate (BBO) crystals. We drive the process with an ultrafast pulsed Ti:Sapphire laser, frequency doubled to 404nm. In the type I case, we use the repeated type I geometry \cite{KwiatRepeatedType1}. We use a beam splitting mask or Gaussian mask (GM) to divide the pump beam into any desired number of `sub-Gaussian' inputs. Each of these sub-Gaussians can in turn generate its own SPDC emission cone.  The configuration of the mask will determine the number and the manner of overlap of the down-converted cones; as illustrated in FIG. \ref{Mask Figure}. 
 The intersection of the SPDC emissions ensures the quantum states of each photon are indistinguishable, which can be extended to multipartite states when the rings each contribute more than one photon. The indistinguishability afforded by this overlap allows us to essentially duplicate [4] the quantum state from each SPDC emission at the intersections, yielding excess entangled photons as compared to the single pump case.
 
 A similar argument can be followed for the type II interaction where two intersection are already produced \cite{KwiatHighIntensityType2}.  Other investigations have shown entangled photon yield \cite{ThreeParticle}. For example, sources with repeated orthogonal type II crystals have generated an increase of entanglement products \cite{MultipleEnt}. Here, we study how the input geometry can be altered to accomplish similar and improved results. We show that the orientation of the Type II crystal also controls the manner of the cones overlap.
 %
\section{\label{sec:level1}Theory}
In SPDC, an incident photon is absorbed by a nonlinear medium to excite the emission of a photon pair. These emitted photons have special correlations, with polarization characteristics having been studied extensively in the past \cite{TheoryParametric, TheoryTwoPhoton, TheoryTwoPhotonPulsed, OAMEnt}. For the purpose of this paper, we will only refer to polarization entanglement; arguments, however, can be made for the extension of entanglement across other degrees of freedom between the overlapping SPDC emission cones (i.e. energy, momentum etc.). In all cases we present here, the down conversion is assumed to be degenerate unless it is otherwise explicitly indicated. 

It will be fruitful to define the Bell states as follows:
\begin{equation}
\begin{split}
   \ket{\phi^{\pm}} &= \frac{1}{\sqrt{2}}\qty[\ket{HH}\pm e^{i\varphi}\ket{VV}], \\
   \ket{\psi^{\pm}} &= \frac{1}{\sqrt{2}}\qty[\ket{HV}\pm e^{i\varphi}\ket{VH}].
\label{BellStates}
\end{split}
\end{equation}
The $\phi$ and $\psi$ Bell states are associated with type I and type II processes respectively, due to their ability to readily prepare photons into the states. The relative phase $\varphi$ in (\ref{BellStates}) is determined by phase matching constraints. When referring to the overall states for a subset of down-converted photons from the intersections of a given mask, we will use capitalized $\Phi$ and $\Psi$ for the total wavefunction from types I and II respectively. This can be described by:
\begin{equation}
\begin{split}
   \ket{\phi^{\pm}} &\xrightarrow{G.M.}\ket{\Phi} \\
   \ket{\psi^{\pm}} &\xrightarrow{G.M.}\ket{\Psi},
\label{Transform}
\end{split}
\end{equation}
where $G.M.$ accounts for the transformation induced by the Gaussian mask.

The system of Bell states are composed of two qubits, thus the usual SPDC process conducted with proper phase compensation creates two entangled qubits. In our approach, the indistinguishability in the emission pattern causes entanglement in a higher-dimensional system, creating entangled qubits in excess of two. 
\begin{figure}
\centering\includegraphics[width=\columnwidth]{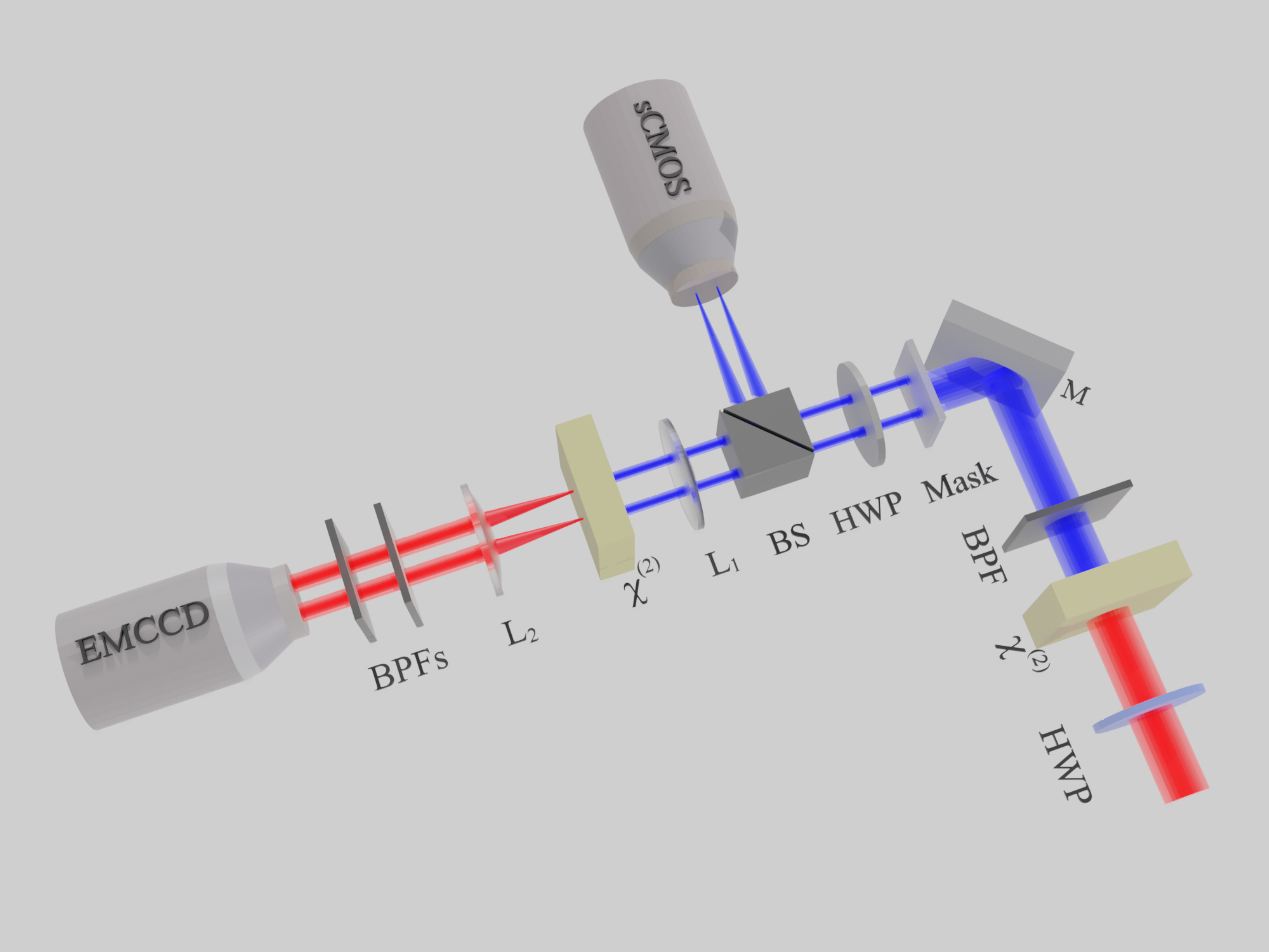}
\caption{Schematic of setup including Half-Wave Plates (HWP), BBOs ($\chi^{(2)}$), Band Pass Filters (BPF), Beam Splitter (BS), Mirrors (M), lenses (L), and our beam masks. The sCMOS and EMCCD image the pump and down converted light respectively.}
\label{Setup}
\end{figure}
\section{\label{sec:level1}Apparatus and Experimental Design}
FIG. \ref{Setup} is an illustration of our experimental setup. The laser source (not shown) is a mode locked regenerative Ti:sapphire laser with repetition rate of 3kHz,  pulse energy of 1.67 mJ, with a 38 femtosecond  pulse duration at 808 nm center wavelength. We use $\beta-Ba(BO_2)_2$ or BBO crystals as our nonlinear media for SPDC production. For type I SPDC we use two adjacent 500 $\mu$m thin nonlinear crystals that are oriented orthogonally such that a vertically or horizontally polarized pump photon can down-convert into a pair of horizontally or vertically polarized photons in the first (second) crystal. For type II SPDC we use a single 1000 $\mu$m thick birefringent BBO crystal. The crystals are pumped by the frequency doubled beam (404 $nm$) via a separate type I BBO crystal as shown in FIG. \ref{Setup}. 

The frequency doubled beam is transformed into sub-Gaussian beams using the GM. The resulting beams are then focused by lens $L_1$ on either the type I or II BBO to produce the desired down-converted cones. The half-wave plate (HWP) placed after  $L_1$ is used to match the crystals' polarization axes. For repeated type I SPDC we set the polarization of the fundamental beam at  45$^{\circ}$, so that the input polarization has equal components along each crystals' optical axis. For the type II process, we match the fundamental polarization to the crystal optical axis. The down-converted light is collected by lens $L_2$ and isolated with two band pass filters (BPF) before being captured by an electron multiplying coupled charged device (EMCCD). An image of the input (pump) beam was recorded with a separate scientific complementary metal oxide semiconductor (sCMOS) camera for each of the masks.

The GM is designed to have 1, 2, 3, or 4 apertures, measuring approximately 2 mm in diameter, and spaced 1.5 mm apart. In the scenario where there are 4 apertures, the spacing refers to the closest neighbors and not the diagonal of the grid. In both the repeated type I and type II processes, the GM geometry can be seen in the emission pattern from the crystal, revealing an increased brightness in the overlapping regions.

\section{\label{sec:level1}Results}

We verify our down conversion through sufficient filtering before imaging it on the EMCCD. We tune the polarization of the pump beam using the HWP to fully extinguish the observed rings. To ensure that none of the rings are reflections, we cover the apertures and separately observe the corresponding emission ring disappear. For simultaneous SPDC processes, the overlap of each emission can be adjusted by the mask design. The single aperture case is given for each process as a reference to the normal SPDC process conducted without a mask.
\begin{figure}
\centering\includegraphics[width=\columnwidth]{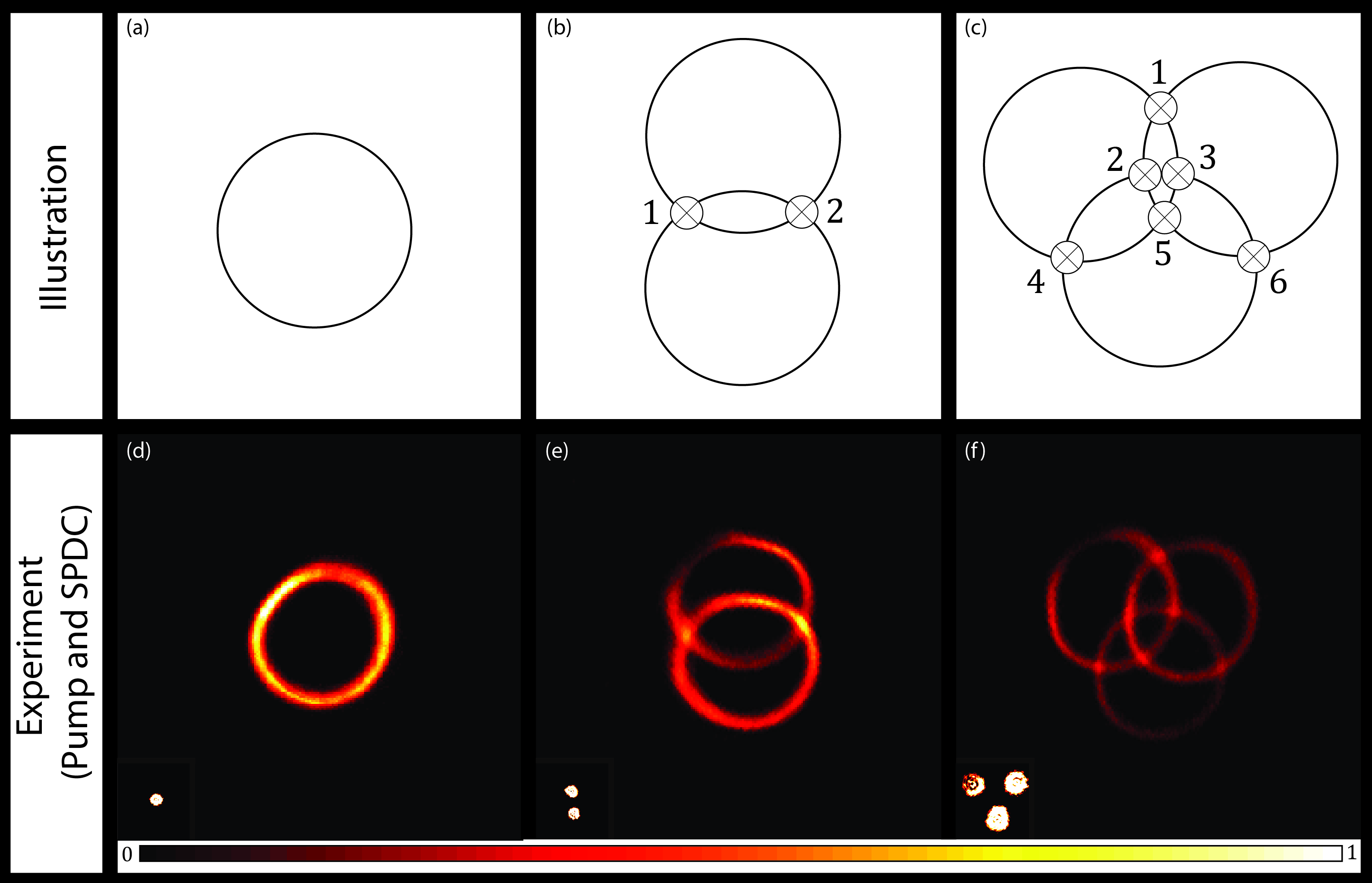}
\caption{(a)-(c) Illustrations of repeated Type I SPDC cones generated by a single, double, and triple "sub-Gaussian" pump, with the desired overlap labeled.  (d)-(f) The correspond experimental results obtained using GMs designed with 1, 2, and 3-apertures. For each case the pump beam was imaged and inset with the SPDC emissions. While there are no intersection in a), the entire ring is entangled due to its origin from both type I BBOs.}
\label{RepType1Figure}
\end{figure}

Illustrations of the SPDC emission cones along with the expected results for repeated type I are displayed in FIG. \ref{RepType1Figure}. The increased brightness at the intersection points of the SPDC cone  confirms the higher photon density present in these overlapping regions. FIG. \ref{RepType1Figure}-(e) shows the 2-aperture GM process conducted with repeated type I. For each aperture, the downconversion process will create entangled pairs of photons in the $\ket{\phi^{+}}$ state. The experimental setup is designed with phase compensation together with rigorous phase matching. This allows for the exponential phase factor in (\ref{BellStates}) to be set to:  $e^{i\varphi} = 1$. We can thus determine one possible form of the wavefunction for the photons emitted at the overlap as follow:
\begin{equation}
\begin{split}
     \ket{\Phi} &= N[\ket{HHHH} + \ket{HHVV} + \ket{VVVV} + \ket{VVHH} \\
     &+ \ket{HVHV} + \ket{HVVH} + \ket{VHVH} + \ket{VHHV}],
     \label{Type 1 Ex}
\end{split}
\end{equation}
where $N=\frac{1}{\sqrt{8}}$. If all outcomes are equally probable, we can use the shorthand $\ket{ij}\otimes\ket{kl}=\ket{ij}\ket{kl}=\ket{ijkl}$ for the outer (tensor) product. The ordering has been done to reveal the factoring admitted by the state. Rearranging gives us:
\begin{equation}
\begin{split}
     \ket{\Phi}  =& \frac{1}{2\sqrt{2}}[\ket{HH}(\ket{HH}+\ket{VV})+\ket{VV}(\ket{VV}+\ket{HH}) \\
     &+ \ket{HV}(\ket{HV}+\ket{VH})+\ket{VH}(\ket{VH}+\ket{HV})], \\
     =& \frac{1}{\sqrt{2}}\qty[\ket{\phi^{\pm}}\ket{\phi^{\pm}} \pm \ket{\psi^{\pm}}\ket{\psi^{\pm}}].
     \label{Type 1 Algebra}
\end{split}
\end{equation}
The addition has been changed to plus-minus in the last line to account for the real values from the exponential phase factors. From this we can see that the source (repeated type I) behaves differently when it is driven with two pumps beams with overlapping emission cones. The resulting state is in an entangled state between products of Bell states, of which now contains the product state $\ket{\psi^{\pm}}\ket{\psi^{\pm}}$.

\begin{figure}
\centering\includegraphics[width=\columnwidth]{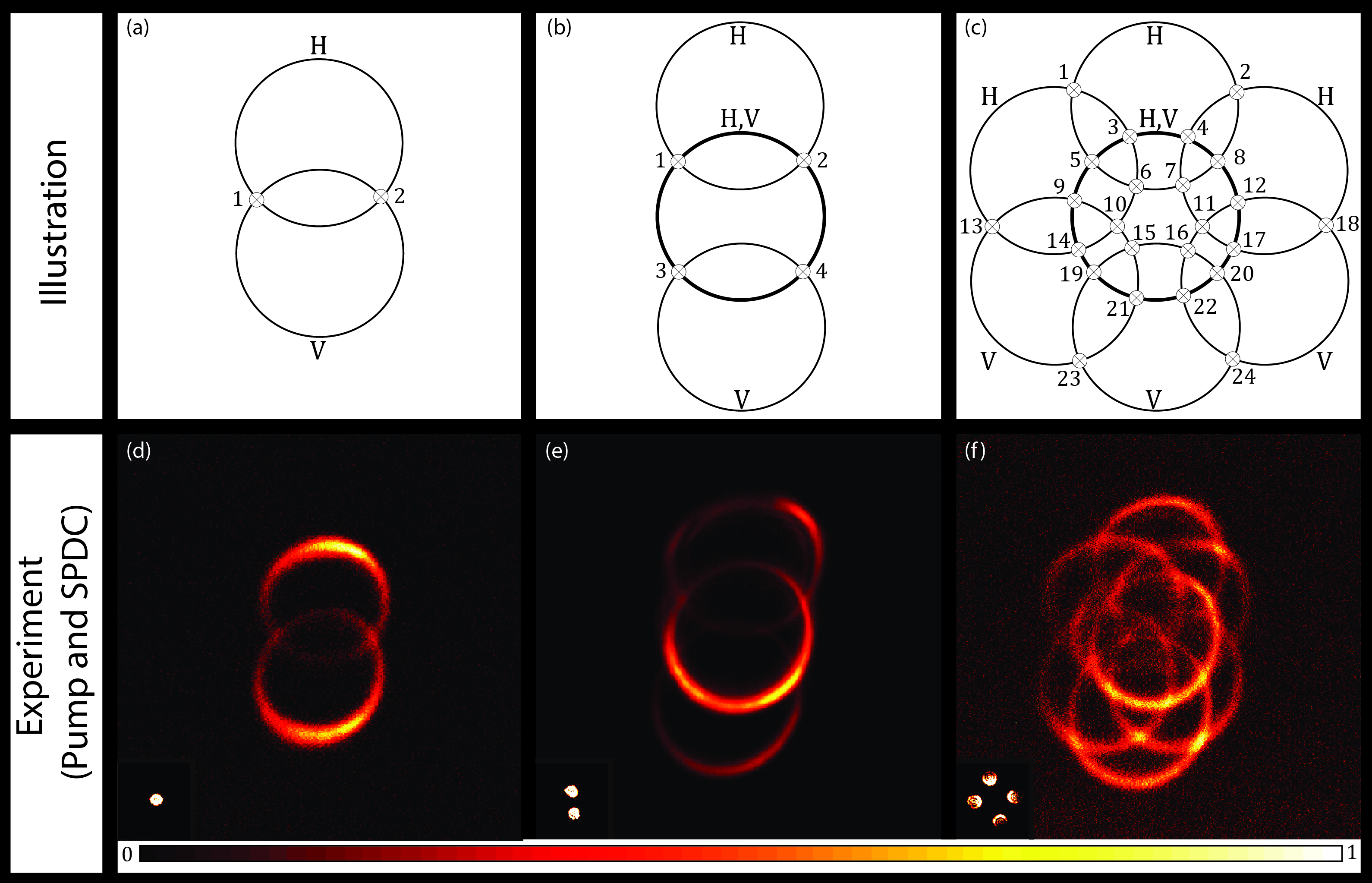}
\caption{(a)-(c) Illustrations of Type II SPDC cones generated by a single, double, and quadruple "sub-Gaussian" pump, with the desired overlap labeled.  (d)-(f) The corresponding experimental results obtained using GMs designed with 1, 2, and 4-apertures. For each case the pump beam was imaged and inset with the SPDC emissions. In (f) the center ring is composed of overlapping rings from the emissions from both the top and bottom apertures.}
\label{Type2Figure}
\end{figure}

For type II, the intersections already afford polarization entanglement in the emitted photons. However, by dividing the beam before passing through the type II crystal, we can completely overlap rings of orthogonal polarization from the separate processes, producing polarization entanglement across the whole overlapped ring, much like in the case of repeated type I, but instead prepared into the $\ket{\psi^{+}}$ Bell states [16]. Furthermore, the photons at the ring intersections from individual SPDC processes will be entangled with each other as distinguishability gets erased upon full spatial overlap. The emission patterns are given along with the expected results in FIG. \ref{Type2Figure}. Type II SPDC is inherently less bright than type I (due to stricter phase matching conditions), thus the background noise is more present in the captured emissions. 

\begin{figure}
\centering\includegraphics[width=\columnwidth]{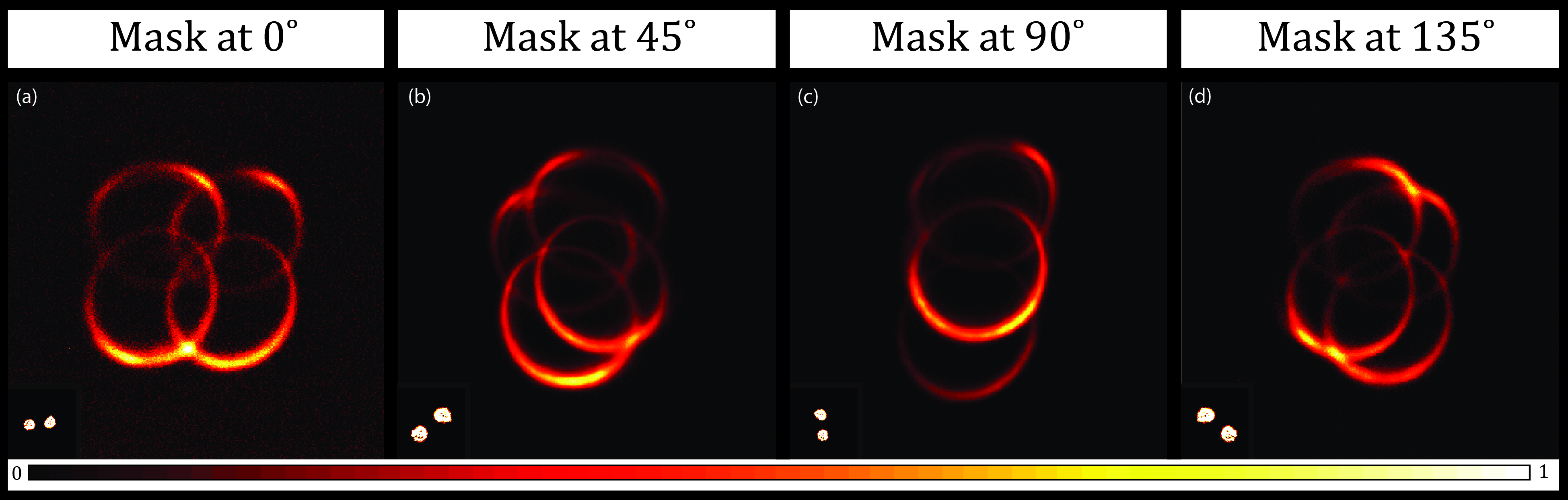}
\caption{Type II SPDC emission with a 2-aperture mask. The rotation of the mask demonstrates the full angular control of the emission.}
\label{Type2Orientations}
\end{figure}

Conducting type II SPDC with a 2-aperture mask is fundamentally different from producing a mixture of type II correlated photons, which we again show with a brief 2-aperture example, shown in FIG. \ref{Type2Figure}-(e). 
\\It should be noted there are many more geometries afforded by conducting the type II process with multiple apertures due to the asymmetry in the emission (with respect to polarization). We only examine one case here, but the process follows similarly for any given emission pattern. Four different orientations of type II SPDC with a 2-aperture mask can be seen in FIG. \ref{Type2Orientations}.

By symmetry arguments, one can see that the wavefunction for the photon states at the overlapping regions includes terms that don't appear when computing the outer product of the wavefunction for a type II process (given in $\ket{\psi^+}$ from eq. (\ref{BellStates})) with itself [16]. The latter gives:
\begin{equation}
\begin{split}
     (\ket{HV} &+ \ket{VH}) \otimes (\ket{HV} + \ket{VH}) \\
     &= \ket{HVHV} + \ket{HVVH} + \ket{VHHV} + \ket{VHVH}, 
     \label{Outer Product}
\end{split}
\end{equation}
where we have ignored the normalization and phase factor. However, we can see that the wavefunction for the overlapping regions should be of the form:
\begin{equation}
\begin{split}
    \ket{\Psi} =& N[\ket{HVHV} + \ket{HVVH} + \ket{VHHV} + \ket{VHVH} \\
    &+\ket{HHVV}], \\
    =& N[2\ket{\psi^{\pm}}\ket{\psi^{\pm}} \pm \ket{HHVV}],
     \label{Type 2 Ex}
\end{split}
\end{equation}
where $N=\frac{1}{\sqrt{5}}$ if all SPDC outcomes are equally probable, and the last term is determined by the emission polarizations (we have chosen $\ket{HHVV}$ instead of $\ket{VVHH}$). Again the addition has been changed to plus-minus to account for the real values of the phase factor. Comparing this with eq. (\ref{Outer Product}), we see a new term. This is due to the fact that the configuration allows the set of intersections above the middle ring to have the same polarization, with the intersections below polarized orthogonally. Neither of these states are possible in the product of two type II emissions because the sets of intersections belonging to each process must posses one vertically and horizontally polarized photon each.

To increase the amount of photons entangled, care must be taken to compensate for the different phases of the emitted light. Accomplishing this leads to an increase in the amount of entangled photon pairs, although it is still dependent on the SPDC configuration and the detection scheme.

\section{Conclusion}

We have shown a novel but simple method to increase the entanglement products from SPDC. We've introduce a Gaussian mask  to divide the pump beam into any desired number of `sub-Gaussian' inputs to induce multiple down-converted emission cones from a single nonlinear crystal. We've shown that the mask design determines the number and the manner of overlap of the down-converted cones. This method can be easily extended to CW laser setups, which would eliminate the need to consider clocking effects of the pulse train. This simple approach has promising potential for photonic qubit production setups, being limited only by the size of the nonlinear crystal and pump beam waist (and by extension the pump power and damage threshold of the nonlinear medium per square area). We believe this approach can help simplify and reduce costs for entanglement generation, and thus greatly contribute to photonic based quantum computing.


\begin{thebibliography}{}

\bibitem{GHZExp}
Z. Zhao, T. Yang, A. Chen, A. N. Zhang, M. Zukowski, and J. W. Pan,
Phys. Rev. Lett. \textbf{91}, 180401 (2003).

\bibitem{QOBook}
G. Jaeger, and A. V. Sergienko,
\uline{Progress in Optics}, vol. 42.
E. Wolf, ed. (Elsevier, 2001), pp.277-324.

\bibitem{KwiatRepeatedType1}
P. G. Kwiat, E. Waks, A. G. White, I. Appelbaum, and P. H. Eberhard,
Phys. Rev. A \textbf{60}, R773-R776 (1999).

\bibitem{NoCloningThm}
No state information is obtained, so this doesn't conflict with the no-cloning theorem

\bibitem{KwiatHighIntensityType2}
P. G. Kwiat, K. Mattle, H. Weinfurter, A. Zeilinger, A. V. Sergienko, and Y. Shih,
Phys. Rev. Lett. \textbf{75}, 4337-4341 (1995).

\bibitem{ThreeParticle}
A. Zeilinger, M. A. Horne, H. Weinfurter, and M. Zukowski,
Phys. Rev. Lett. \textbf{78}, 3031-3034 (1997).

\bibitem{MultipleEnt}
M. L. Fanto, R. K. Erdmann, P. M. Alsing, C. J. Peters, and E. J. Galvez,
in \uline{Quantum Information and Computation IX}, vol. 8057 
E. Donkor, A. R.Pirich, and H. E. Brandt, eds., International Society for Optics and Photonics, (SPIE, 2011), pp. 41-52.

\bibitem{TheoryParametric}
C. K. Hong and L. Mandel,
Phys. Rev A \textbf{31}, 2409-2418 (1985).

\bibitem{TheoryTwoPhoton}
M. H. Rubin, D. N. Klyshko, Y. H. Shih, and A. V. Sergienko, 
Phys. Rev. A \textbf{50}, 5122-5133 (1994).

\bibitem{TheoryTwoPhotonPulsed}
T. E. Keller, and M. H. Rubin,
Phys. Rev. A \textbf{56}, 1534-1541 (1997).

\bibitem{OAMEnt}
S. Franke-Arnold, S. M. Barnett, M. J. Padgett, and L. Allen,
Phys. Rev. A \textbf{65}, 033823 (2002).


\bibitem{PixelEnt}
M. N. O'Sullivan-Hale, I. Ali Khan, R. W. Boyd, and J. C. Howell,
Phys. Rev. Lett. \textbf{94}, 220501 (2005).

\bibitem{MacroscopicRegions}
A. Burlakow, M. Chekhova, D. Klyshko, S. Kulik, A. Penin, Y. Shih, and D. Strekalov,
Phys. Rev. A \textbf{56} (1997).

\bibitem{Horodecki}
R. Horodecki, P. Horodecki, M. Horodecki, and K. Horodecki,
Rev. Mod. Phys. \textbf{81}, 865-942 (2009).

\bibitem{Blackhole}
P. Levay,
Phys. Rev. D \textbf{74}, 024030 (2006).

\bibitem{Phase}
Setting the exponential phase factor to +1

\end{thebibliography}

\end{document}